\documentclass[twoside]{article}
\usepackage{fleqn,espcrc2}

\usepackage{graphicx}

\newcommand{\AmS}{{\protect\the\textfont2
  A\kern-.1667em\lower.5ex\hbox{M}\kern-.125emS}}

\title{Logarithmically rising $hh$ and $\gamma h$ cross-sections~:\\
          some features of model versus experimental data.}

\author{E. Martynov \address[BITP]{Bogoliubov Institute for Theoretical
Physics, National Academy of Sciences of Ukraine,\\ 03143, Kiev-143,
Metrologicheskaja 14b, Ukraine\\
e-mail: martynov@bitp.kiev.ua }}

\begin{document}

\begin{abstract}
A Regge model of the Pomeron with intercept equal to one, leading to rising
cross-sections is considered. Analysis of the experimental data on hadron
and photon induced interactions is performed within the model. It is shown
that the available pure hadronic data as well as data on DIS are
compatible with a high energy logarithmic behaviour of cross-sections and
do not require a Pomeron intercept above one. The very important r\^ole
of the preasymptotic contributions is especially emphasized. \vspace{1pc}
\end{abstract}

\maketitle

In this talk I would like to discuss the importance of preasymptotic terms
when the available data are analyzed and described in the
framework of Regge approach. These terms strongly influence the
conclusions about properties of the asymptotic term, the Pomeron.
I shall explicit as an example
the soft Pomeron model, not violating the Froissart-Martin bound, in which
the contribution of secondary Reggeons, in particular $f$ Reggeon, can not
be neglected, even at the highest accelerator energies. The model
describes the hadron elastic scattering as well as the proton
structure function.

The elastic scattering amplitude in Regge approach is given by the sum of
Pomeron and secondary Reggeons
\vspace{-0.2cm}
\begin{eqnarray*}
A(s,t)&=&P(s,t)+R(s,t)\\
&\equiv&P(s,t)+f(s,t) \ + \ [\pm \omega \pm\rho \pm a_{2},... ]
\end{eqnarray*}

\vspace{-0.2cm} \noindent
 where the signs of terms in square brackets
depend on the process under consideration.

I concentrate here firstly on two aspects~: the exchange
degeneracy of various Reggeons and the phenomenological separation be\-tween the
Pomeron and the $f$ Reggeon.

$\bullet$ {\bf Exchange degeneracy.} The hypothesis is based on the fact
that $f, \omega , \rho , a_{2}$ trajectories seem to coincide on a
Chew-Frautschi plot.

The linear parameterization \vspace{-0.2cm}
\begin{eqnarray*}
  \alpha_{e-d}(m^{2})=\alpha_{e-d}(0)+\alpha'_{e-d}m^{2}
\end{eqnarray*}

\vspace{-0.2cm} \noindent of the single exchange-degenerate trajectory,
fitted to the data, gives \cite{dgmp} \vspace{-0.2cm}
\begin{eqnarray*}
  \alpha_{e-d}(0)=0.449, \qquad \alpha'_{e-d}=0.901\ {\rm GeV}^{-2}
\end{eqnarray*}

\vspace{-0.2cm} \noindent
and an extremely high $\chi^{2}$, namely
$\chi^{2}/dof\approx 118$.

Fitting in isolation each of the trajectories, we obtain the
result shown in Fig.~1. Numerically we have \vspace{-0.2cm}
$$ \begin{array}{ll}
    \alpha_{f}(0)=0.697\pm0.041,&\\  \alpha'_{f}=(0.801\pm0.002)\
    {\rm GeV}^{-2},  & \!\! \chi^{2}/dof=6.01, \\
    \alpha_{\omega}(0)=0.436, &\\ \alpha'_{\omega}=0.923\ {\rm GeV}^{-2},
    &   \mbox{(not fitted)},\\
    \alpha_{\rho}(0)=0.478\pm0.001, &\\ \alpha'_{\rho}=(0.880\pm0.002)\
    {\rm GeV}^{-2},  &\!\! \chi^{2}/dof=3.31, \\
    \alpha_{a_{2}}(0)=0.512\pm0.041, &\\ \alpha'_{a_{2}}=(0.857\pm0.023)\
    {\rm GeV}^{-2},  &\!\! \chi^{2}/dof=0.42.
  \end{array}
$$
\noindent
The bad $\chi^{2}$ for the $f$ trajectory is due to its
evident nonlinearity (see below).

Thus the first conclusion is the following~:

\noindent
 {\bf \it the available data on mesons lying on the $f$, $\omega$, $\rho$
 and $a_2$ Regge trajectories contradict the exchange degeneracy assumption.}

It should be noted that, in accordance with the conclusion derived in
\cite{dgmp,CEKLT}~\footnote{Because of the very restricted size of the
talk I give only the references to our papers where further references can be
found.}, the hypothesis of exchange degenerate
trajectories is not supported also by the forward scattering data.

\begin{center}
\includegraphics[scale=0.35]{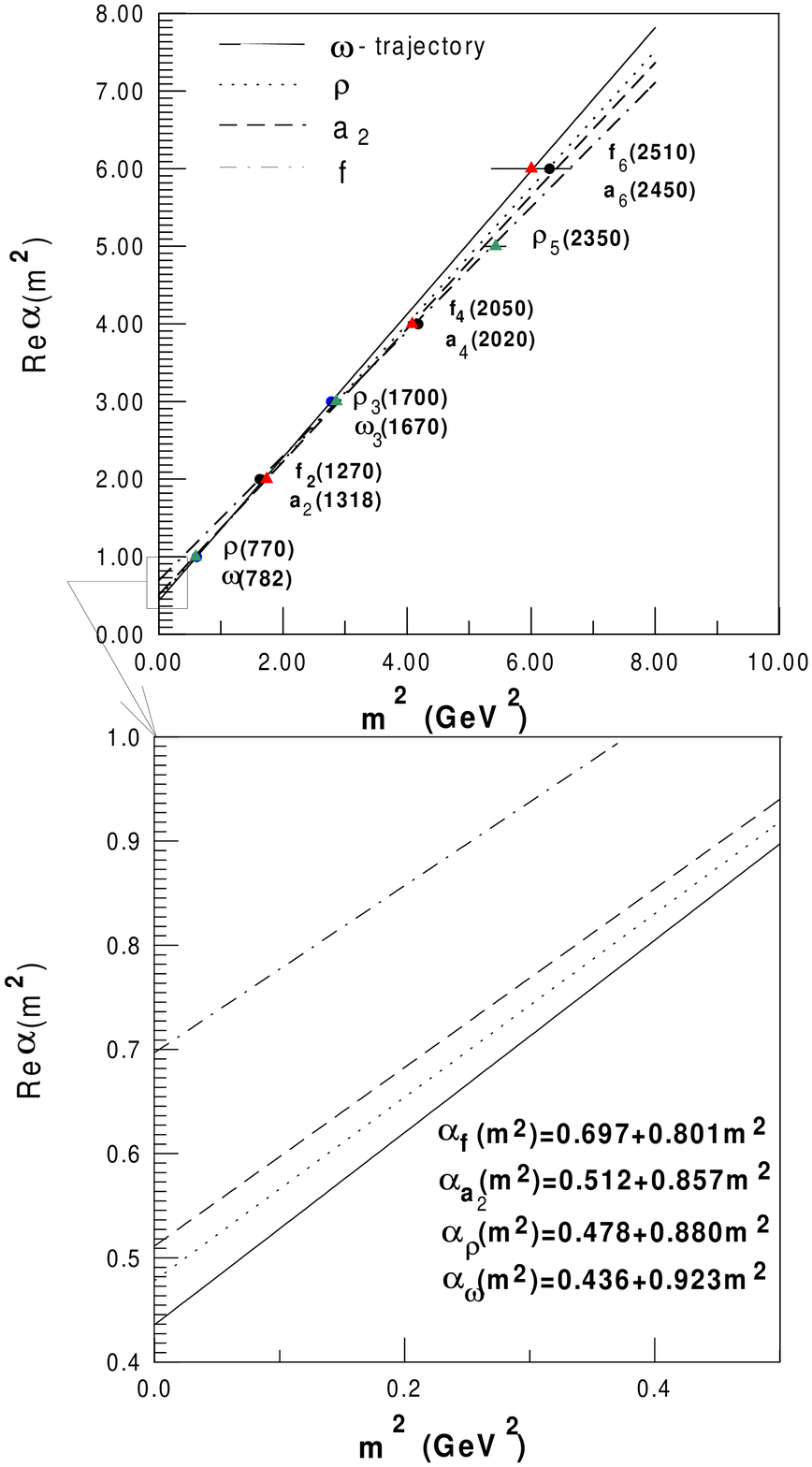}
\end{center}
{\bf Fig.1} {\small Chew-Frautschi plot for $f$, $\omega$, $\rho$ and
$a_2$ Regge trajectories taken separately assuming linearity (the figure
below is an enlargement for small masses).}

\medskip

$\bullet $ {\bf $f$ trajectory.} The nonlinearity of the $f$ trajec\-tory is
illustrated unambiguously in Fig.~2, where the above mentioned linear
fit is shown together with the parabola passing exactly through the three known
resonances.

\begin{center}\label{realf}
\includegraphics*[scale=0.25]{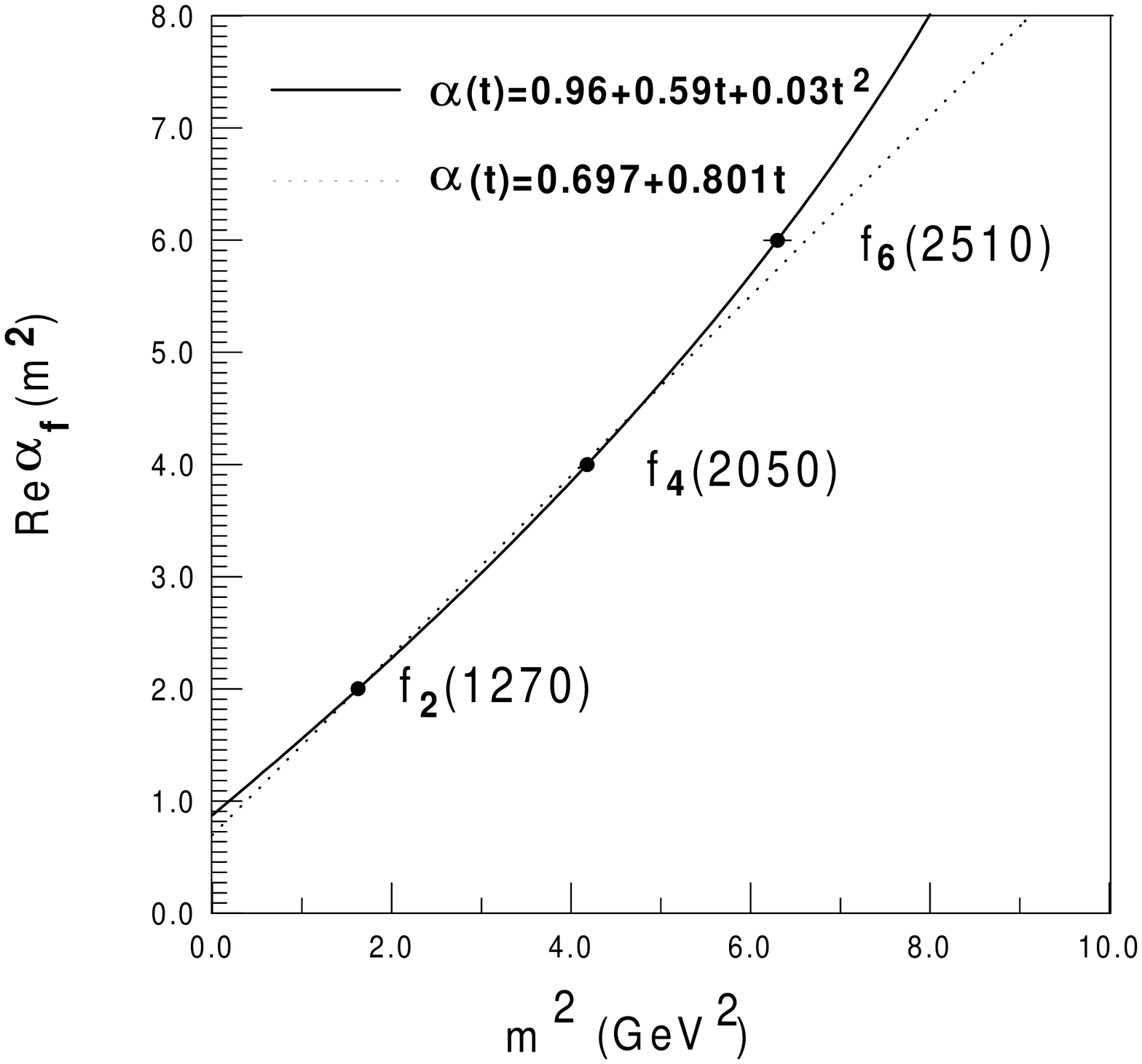}
\end{center}
{\bf Fig.2} {\small Real part of the $f$ trajectory versus the squared
mass of the re\-so\-nan\-ce, $m^2$. Solid line is the parabola passed
through the known resonances. Dashed straight line is the result of a
linear fit.}

\medskip
Certainly, trajectories must rise slower than first power of $t$ at $t\to
\infty$. A more realistic $f$~tra\-jectory chosen for example in the form
 \vspace{-0.1cm}
\begin{eqnarray*}
 \alpha_f(t)=\alpha_f(0)&+&\beta_1(\sqrt{4m_{\pi}^2}-\sqrt{4m_{\pi}^2-t}\
)\\&+&\beta_2(\sqrt{t_1}-\sqrt{t_1-t}\ )
\end{eqnarray*}

\vspace{-0.1cm}
\noindent leads to $0.77<\alpha_f(0)<0.87$.

Thus, we obtain a very interesting and important phenomenological
consequence of the nonlinearity~: a higher intercept of the $f$ trajectory.
Coming to the next conclusion~:

\noindent {\it the intercept of the $f$ trajectory is $\ge 0.7$,
but most probably, the lower bound is larger than this value.}

$\bullet$ {\bf Separation of Pomeron and $f$ Reggeon.} Generally, there is
an evident correlation between the intercept of the $f$ Reggeon and the
model for the Pomeron. This is due to the fact that in all known processes
Pomeron and $f$ Reggeon contribute additively. As a rule, a higher $f$
intercept is associated with a slower growth with energy due to the Pomeron
contribution~\cite{dgm}. In Fig.~3 we illustrate this observation and show
how $\alpha_f(0)$ is correlated with a power of $\ell n s$ in the
behaviour of the total cross-section, if the forward $pp (\bar pp)$
scattering amplitudes are parametrized in the form
\vspace{-0.3cm}
\begin{eqnarray*}
  A^{(hh)}(s,0)&=&i\Big[C_1+C_2\ell n^{\gamma}(-is/s_0)\Big]\\
  &+&f(s,0)\pm \omega (s,0),
\end{eqnarray*}
with the standard form of secondary Reggeons contribution.

\begin{center}
\includegraphics*[scale=0.3]{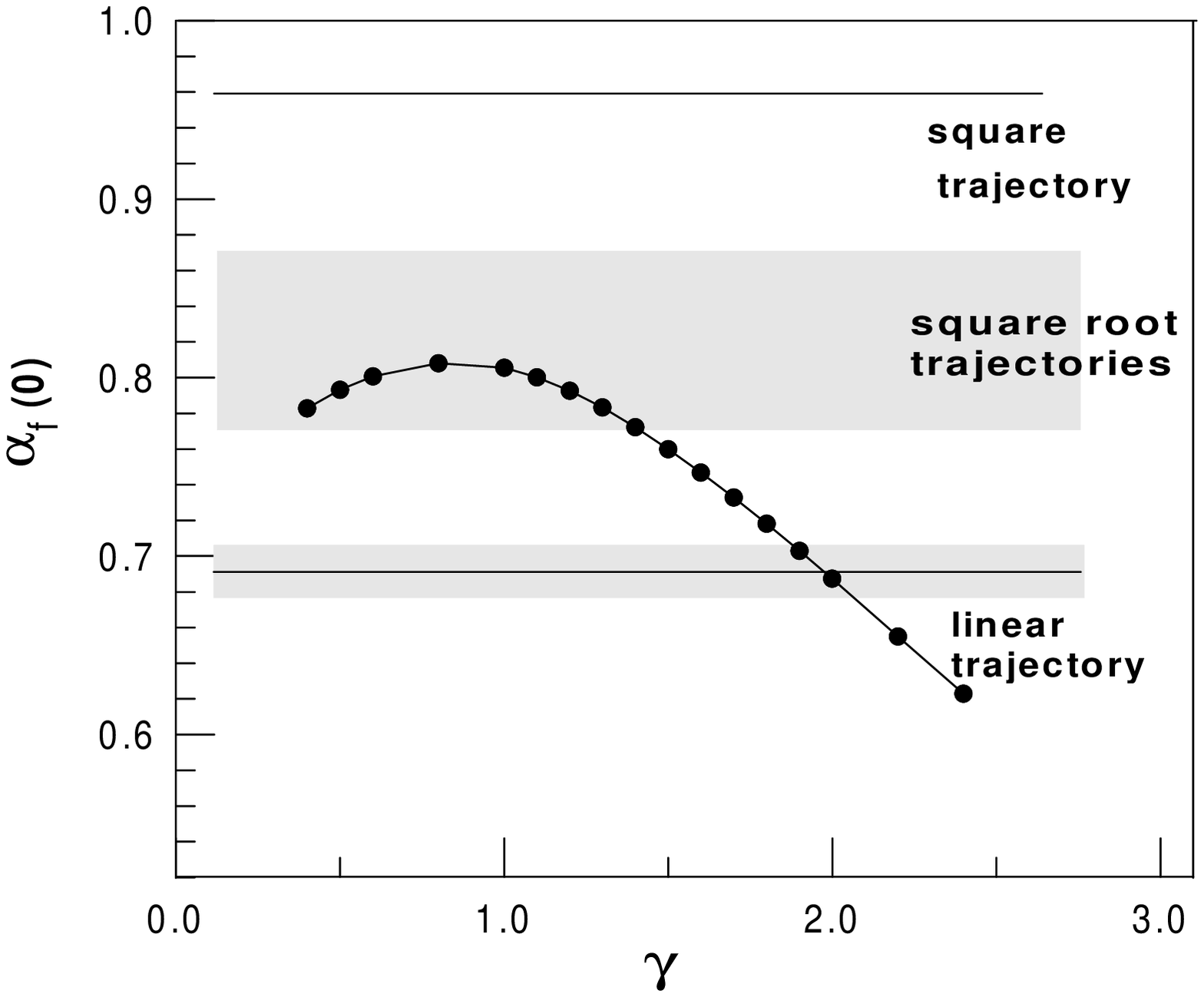}
\end{center}

{\bf Fig.3} {\small Intercept of the $f$ trajectory correlated with the
power $\gamma$ of $\ell n s$ in $\sigma_{tot}(s)$. Intervals for intercept
of some trajectory parameterizations are shown.}

\medskip
Taking into account the restriction $\alpha_{f}(0)>0.7$, revealed from
spectroscopy data, one can conclude that $\gamma <2$ or that total
cross-section rises slower than $\ell n^{2}s$.

On the other hand it is known~\cite{dgmp,CEKLT,dgm,dglm,maqm} from a
comparison of various Pomeron models that the "best" description (or
$\chi^{2}$) is obtained if the Pomeron contribution to the total
cross-section has the form
$\sigma_{P}=C_{1}+C_{2}\ell n(s/s_{0})$. The difference in $\chi^{2}$ is
quite small ($\sim$ a few percents) if $\sqrt{s}\geq \sqrt{s_{min}}=9$ GeV.
An additional advantage of the logarithmic model is its
stability (of the parameters as well as the $\chi^{2}$) when
$\sqrt{s_{min}}$ is decreasing down to 5 GeV.

It is interesting to note two points concerning the logarithmic model~:

\noindent
1.Intercept of the $f$ Reggeon takes approximately its maximal value of
Fig.~3, $\alpha_{f}(0)\approx 0.8$.

\noindent
2.If the well known Supercritical Pomeron
$
 \widetilde{P(s,0)}=iC(-i\frac{s}{s_{0}})^{\Delta}
$
is generalized for two components
$
i[\tilde C_{1}+\tilde C_{2}(-is/s_{0})^{\Delta}]
$ (provided the Regge trajectories are non exchange degenerate) then a fit
to the total cross-sections leads to a very small value of $\Delta
(\approx 0.001)$. Hence, one can approximate
 \begin{eqnarray*}
\tilde C_{1}+\tilde C_{2}(-is/s_{0})^{\Delta}\approx C_{1}+C_{2}\ell
n(-is/s_{0}),\\
C_{1}=\tilde C_{1}+\tilde C_{2},\quad C_{2}=\Delta \tilde C_{2}
\end{eqnarray*}
with all parameters ($C_{1},C_{2}$ as well as couplings and intercepts of
the secondary Reggeons) coinciding with those obtained in the logarithmic
Pomeron model.

This observation is valid not only for $pp$ and $\bar pp$ cross-sections
but also for all hadronic, $\gamma p$ and $\gamma \gamma$ cross-sections.

It is necessary to emphasize that in this model the preasymptotic $f$ term
gives  quite a large part of the total cross-section even at the Tevatron
energy, $\sigma_{f}(\sqrt{s}=1.8 TeV)\approx 3$ mb. It cannot be
neglected at the available energies.

The third conclusion follows~:

\noindent {\it the available data on the total cross-sections of hadron
and photon induced processes are better described in the model yielding
a moderate (logarithmic) rise of the cross-sections.}

\medskip
$\bullet$ {\bf Dipole Pomeron model and elastic scattering.} I present here,
without any details and references for an history of the subject, that can
be found in~\cite{maqm,dgm2}, only the final results concerning the
differential cross-sections of elastic $pp$ and $\bar pp$ interactions in
the so-called Modified Additive Quark Model (MAQM).

In the MAQM, we have assumed that the Pomeron can be coupled
not only with a single quark but also with a pair of quarks,
giving rise to a small
($\sim 10\%$) but important correction. Also, some counting
rules for the Reggeons contribution based on their quark contents have been
suggested.

The above mentioned logarithmic Pomeron model can be named as "Dipole
Pomeron" (DP) model because in the complex momentum plane ($j$-plane) a
double pole at $j=\alpha_{P}(t)$ with $\alpha_{P}(0)=1$ is the dominating
contribution in the amplitude and leads asymptotically at $s\to \infty$ to
$\sigma_{tot}\propto \ell n(s)$.

Applying the MAQM and DP (adding at $t\neq 0$ an Odderon contribution), we
obtain a quite good description (with $\chi^{2}/dof \approx 2.38$) of the
total cross-sections  for $pp$,
$\bar pp$, $\pi^{\pm} p$, $\gamma p$ and $\gamma \gamma $ and of the
differential cross-sections of $pp$ and $\bar pp$
elastic scattering in a wide kinematical region ($\sqrt{s}\geq $5 GeV
for $t=0$~\cite{maqm} and $\sqrt{s}\geq $19 GeV for
$0<|t|\leq 14$ GeV$^{2}$~\cite{dgm2}).

Thus~:

\noindent {\it not only the forward scattering data but also the elastic
scattering at small and large $|t|$ can be described with a high quality
in the Dipole Pomeron model with the intercept $\alpha_P(0)$ equal to one.}

$\bullet$ {\bf Dipole Pomeron model and deep inelastic scattering.}
Another example demonstrating the importance of preasymptotic terms, when
the properties of Pomeron are derived from the experimental data, is the
Dipole Pomeron model for the forward $\gamma^{*} p$ amplitude.

Defining the Dipole Pomeron model for DIS, we start from the expression
connecting the transverse cross-section of $\gamma^*p$ interaction to the
proton structure function $F_2$ and the optical theorem for forward
scattering amplitude
 \begin{eqnarray*}
\sigma_{T}^{\gamma^*p}(W,Q^{2})=\Im m A(W^2,Q^2;t=0)\\
=\frac{4\pi^2\alpha}{Q^2(1-x)}(1+4m_p^2x^2/Q^2)F_2(x,Q^2),
\end{eqnarray*}
where $\sigma_L^{\gamma^*p}=0$ is assumed. The forward scattering at
$W^2=Q^2(1/x-1)+m_p^2$ being far from the threshold $W_{th}=m_p$ is
dominated by the Pomeron and the $f$ Reggeon
 \footnote{We ignored an $a_2$ Reggeon considering the
$f$ term as an "effective" one at $W\geq 3$ GeV.}
\begin{eqnarray*}
&&A(W^2,Q^2;t=0)=P+f,\quad P=P_1+P_2, \\
&&P_1=iG_1(Q^2)\ln(\frac{-iW^2}{m_p^2})(1-x)^{B_1}, \\
&&P_2=iG_2(Q^2)(1-x)^{B_2}. \\
&&f=iG_f(Q^2)(\frac{-iW^2}{m_p^2})^{\alpha_f(0)-1}(1-x)^{B_f}.
\end{eqnarray*}
It evidently follows from the experimental data that
$Q^2\sigma^{\gamma^*p}(W,Q^{2})$ decreases with $Q^2$, at least at high
$Q^2$. We choose
\begin{eqnarray*}
G_i(Q^2)=\frac{g_i}{(1+Q^2/Q_i^2)^{D_i}},
\end{eqnarray*}
expecting $D_i>1$ at high $Q^2$. As it follows from the fit, $D_i$ and
$B_i$ should be functions of $Q^2$. The details of the parameterization of
the real functions $D_i(Q^2), B_i(Q^2)$ are given in~\cite{dlm}.

A fit to the 1389 experimental points in the region $W\geq 3$
GeV$^2$, $0\leq x\leq 0.85$, $Q^2\geq 0$ was performed and a quite good
description of data was obtained~: $\chi^2/dof=1.07$.

Again, as for the pure hadron case,
\noindent
 {\it the preasymptotic contributions in $\sigma^{\gamma^{*}p}$,
the constant (negative) component of the Pomeron term as well as the $f$ term
are very important in the whole considered kinematical region.}

To illustrate this statement, let us consider the "experimental" data on the
"effective" Pomeron intercept and its dependence on $Q^{2}$. This
quantity, $\alpha_{eff}=1+\Delta_{eff}$ (or $1+\lambda_{eff}$) is
extracted from the data on $F_{2}$ in accordance with the parametrization
$F_{2}(x,Q^{2})=C(1/x)^{\lambda_{eff}}$. Strictly speaking a more correct
definition of an effective intercept (or $x$-slope of the structure
function) is as follows
\begin{eqnarray*}
F_{2}(x,Q^{2})=G(Q^{2})(1/x)^{\Delta_{eff}(x,Q^{2})},\\
\Delta_{eff}(x,Q^{2})=\partial F_{2}/\partial \ell n(1/x).
\end{eqnarray*}
$\Delta_{eff}(x,Q^{2})$ coincides with $\lambda_{eff}$ only if it does not
depend on $x$. For an accurate comparison of a model with experiment we
need the data on the local $x$-slope\footnote{An attempt to extract such a
local slope as a function of $x$ and $Q^{2}$ is given in \cite{dlm2}.}
rather than $\lambda_{eff}$ averaged in wide intervals of $x$.
The investigated Dipole Pomeron has an intercept exactly equal to one.
Nevertheless,
due to interferences Pomeron-$f$ Reggeon, the model
is able to describe well the data on averaged $x$-slope as it is shown in
Fig.~4.

\begin{center}
\includegraphics[scale=0.3]{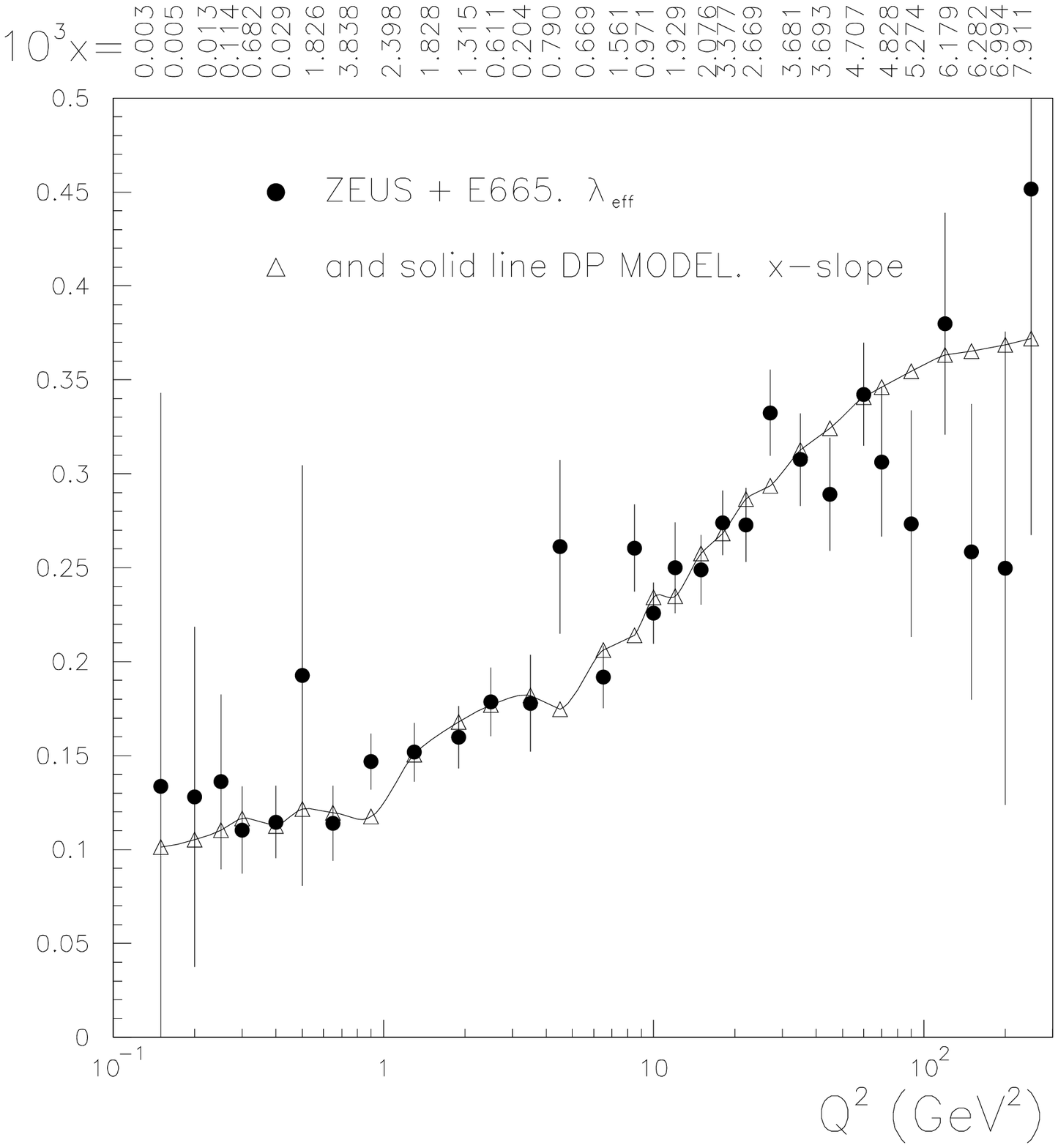}
\end{center}
{\bf Fig.~4} {\small The effective intercept calculated in the Dipole
Pomeron model \cite{dlm} (triangles and solid line) compared with the
experimental data.}
\medskip

We have shown that a careful account of the preasymptotic contributions
allows to describe the available data within the model that does not
violate the asymptotic bounds on the cross-sections.

\end{document}